\documentclass[twoside]{dis04}
\def\runauthor{A. Szczurek}
\def\shorttitle{Kwieci\'nski-CCFM - applications}
\begin{document}

\title{KWIECI\'NSKI-CCFM \\
UNINTEGRATED PARTON DISTRIBUTIONS - \\
A FEW APPLICATIONS
}

\author{Antoni Szczurek}

\address{Institute of Nuclear Physics\\
ul. Radzikowskiego 152, PL-31342 Cracow, Poland\\
and University of Rzesz\'ow\\
PL-35-959 Rzesz\'ow, Poland\\
E-mail: antoni.szczurek@ifj.edu.pl }

\maketitle

\abstracts{A few applications of recent unintegrated
parton distributions from the solution of the recent
equations formulated by Jan Kwieci\'nski are shown.
}

\section{Introduction}

This talk shortly reviews practical applications of the unintegrated
parton distributions which fulfil the Kwieci\'nski evolution equations
\cite{CCFM_b1,CCFM_b2,GKB03,GK03}. The formal aspect is discussed
in a parallel talk by Broniowski. I present some examples of
application of the formalism. This talk is based on
Refs.\cite{KS04,LS04,SC05} where more details can be found.

\section{Production of gauge bosons}

In the formalism of unintegrated parton distributions the nonzero
transverse momenta of gauge bosons are obtained already
in the leading order. The invariant cross section for inclusive
gauge boson production reads then as
\begin{eqnarray}
&\frac{d \sigma}{d y d^2 p_{t,W}} =
\sigma_0^W \sum_{qq'} |V_{qq'}|^2 \int
 \frac{d^2\kappa_1}{\pi} \frac{d^2\kappa_2}{\pi}
\;
\delta^2(\vec{p}_t-\vec{\kappa}_1-\vec{\kappa}_2)
\nonumber\\
&\left[ 
f_{q/1}(x_1,\kappa_1^2,\mu^2) f_{{\bar q'}/2}(x_2,\kappa_2^2,\mu^2) +
f_{{\bar q'}/1}(x_1,\kappa_1^2,\mu^2) f_{q/2}(x_2,\kappa_2^2,\mu^2)
\right] \; .
\label{uPDFs_LO}
\end{eqnarray}
In the equation above the delta function assures the conservation
of transverse momenta in the $q \bar q'$ fusion subprocess.
The momentum fractions are calculated as
$x_{1,2} = \frac{m_{t,W}}{\sqrt{s}} \exp(\pm y)$, where in contrast
to the collinear case $M_W$ is replaced by the transverse mass
$m_{t,W}$.

Introducing unintegrated parton distributions in the space
conjugated to the transverse momenta \cite{CCFM_b1}
\begin{equation}
f_q(x,\kappa^2,\mu^2) = \frac{1}{2 \pi}
 \int  \exp \left( i \vec{\kappa} \vec{b} \right) \; 
\tilde f_q(x,b,\mu^2) \; d^2 b \; ,
\label{Fourier transform}
\end{equation}
and taking the exponential representation of the $\delta$ function
\cite{KS04}
the formula (\ref{uPDFs_LO}) can be written in the equivalent way
\begin{eqnarray}
&\frac{d \sigma}{d y d^2 p_{t,W}} =
\sigma_0^W/\pi^2 \sum_{qq'} |V_{qq'}|^2
\int d^2 b \; J_0(p_t b) \nonumber \\
&\left[ 
\tilde{f}_{q/1}(x_1,b,\mu^2) \tilde{f}_{{\bar q'}/2}(x_2,b,\mu^2) +
\tilde{f}_{{\bar q'}/1}(x_1,b,\mu^2) \tilde{f}_{q/2}(x_2,b,\mu^2)
\right] \; .
\label{b-space_formula}
\end{eqnarray}
In the formulae for $Z^0$ boson production 
$|V_{q q'}|^2$ is replaced by
$\delta_{q q'} \frac{1}{2} (V_q^2 + A_q^2)$.

As already mentioned in the introduction, it is our intention here
to use uPDFs $\tilde{f}_q^{CCFM}(x,b,\mu^2)$ which fulfil b-space
CCFM equations \cite{CCFM_b1,CCFM_b2}.
However, the perturbative solutions $\tilde{f}_q^{CCFM}(x,b,\mu^2)$ do not
include nonperturbative effects such as, for instance,
intrinsic momentum distribution of partons in colliding hadrons.
In order to include such effects we propose to modify the perturbative
solution $\tilde{f}_q^{CCFM}(x,b,\mu^2)$
and write the modified parton distributions
$\tilde{f}_q(x,b,\mu^2)$ in the simple factorized form
\begin{equation}
\tilde{f}_q(x,b,\mu^2) = \tilde{f}_q^{CCFM}(x,b,\mu^2)
 \cdot F_q^{NP}(b) \; .
\label{modified_uPDFs}
\end{equation}
In the present study we shall use two different functional
forms for the form factor
\begin{equation}
F_q^{NP}(b) = F^{NP}(b) = \exp\left(- \frac{b^2}{4 b_0^2}\right) \;
or \; \exp \left( - \frac{b}{b_e} \right)
\label{formfactor}
\end{equation}
identical for all species of partons.
In Eq.(\ref{formfactor}) $b_0$ (or $b_e$) is the only free parameter.
In the next section we try to adjust this parameter to
the experimental data on transverse momentum distribution of $W^{\pm}$.

\begin{figure}[htb] 
    \includegraphics[width=5.0cm]{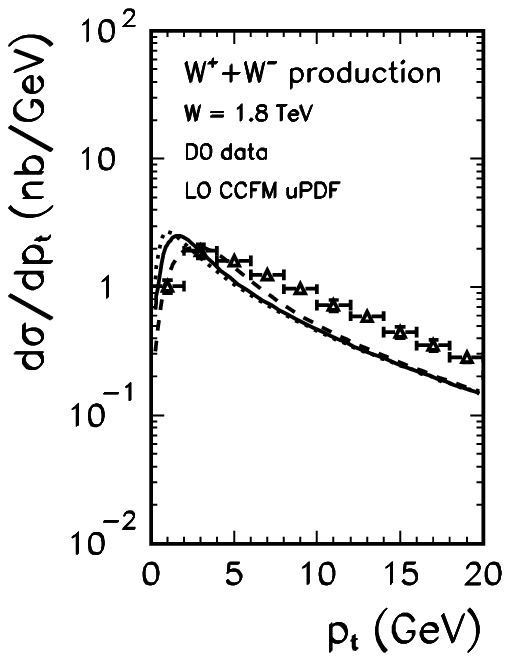}
    \includegraphics[width=5.0cm]{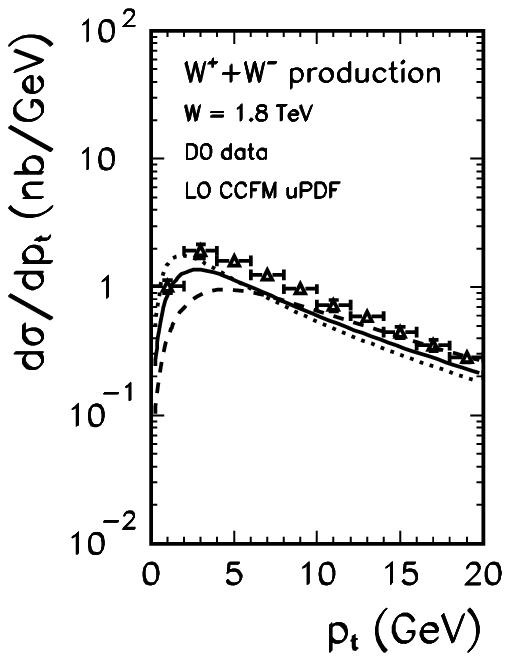}
\caption[*]{
Transverse momentum distribution of $W^+ + W^-$ at W = 1.8 TeV
within the formalism of the CCFM uPDF's with Gaussian (left panel)
and exponential (right panel) form factor.
The three curves correspond to different values of the form factor
parameter $b_0$ or $b_e$ as explained in the text. The experimental
data are taken from Ref.\cite{D0_WZ_2001}.
\label{fig:D0}
}
\end{figure}

As an example in Fig.\ref{fig:D0} we show transverse momentum distribution
(integrated over rapidities) of $W^{\pm}$ in proton-antiproton collissions
at Fermilab at W = 1.8 TeV 
for Gaussian (left panel) and exponential (right panel) form factor.
The three curves in the left panel show results obtained
with different values of the Gaussian form factor parameter
$b_0$: $b_0$ = 0.5 GeV$^{-1}$ (dashed), $b_0$ = 1.0 GeV$^{-1}$
(solid) and $b_0$ = 2.0 GeV$^{-1}$ (dotted).
Similarly, the curves in the right panel
correspond to $b_e$ = 0.5 GeV$^{-1}$ (dashed), $b_e$ = 1.0 GeV$^{-1}$
(solid) and $b_e$ = 2.0 GeV$^{-1}$ (dotted).
The results are overimposed on
the D0 collaboration data \cite{D0_WZ_2001} measured at Fermilab.
The figure clearly demonstrates the importance of the nonperturbative
effects. A better agreement is obtained with the exponential form factor.
In Ref.\cite{KS04} the results of the uPDF approach are compared with
the results of the standard resummation approach.

\section{Charm-anticharm correlations}

The total cross section for quark-antiquark production
in the reaction $\gamma + p \to Q + \bar Q + X$ can be written as
\begin{equation}
\sigma^{\gamma p \to Q \bar Q}(W) =
\int d \phi \int dp_{1,t}^2 \int dp_{2,t}^2 \int dz \;
\frac{f_g(x_g,\kappa^2)}{\kappa^4} \cdot
\tilde{\sigma}(W,\vec{p}_{1,t},\vec{p}_{2,t},z) \; .
\label{master_formula}
\end{equation}
In the formula above $f_g(x,\kappa^2)$ is the unintegrated gluon
distribution.
The gluon transverse momentum is related to the quark/antiquark
transverse momenta $\vec{p}_{1,t}$ and $\vec{p}_{2,t}$ as:
\begin{equation}
\kappa^2 = p_{1,t}^2 + p_{2,t}^2 + 2 p_{1,t} p_{2,t} cos\phi \; .
\end{equation}
%

\begin{figure}[htb] 
    \includegraphics[width=5.0cm]{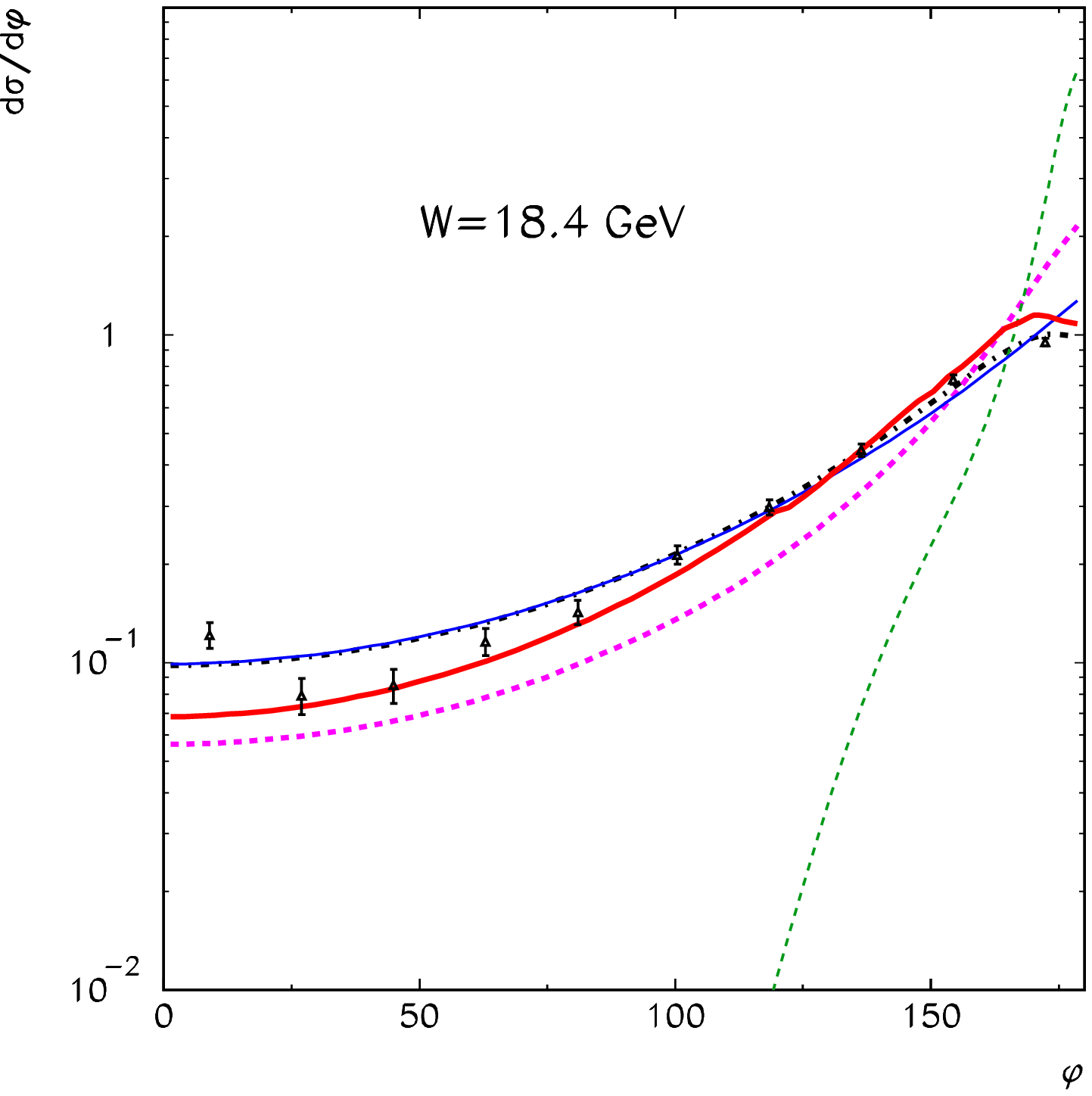}
    \includegraphics[width=5.0cm]{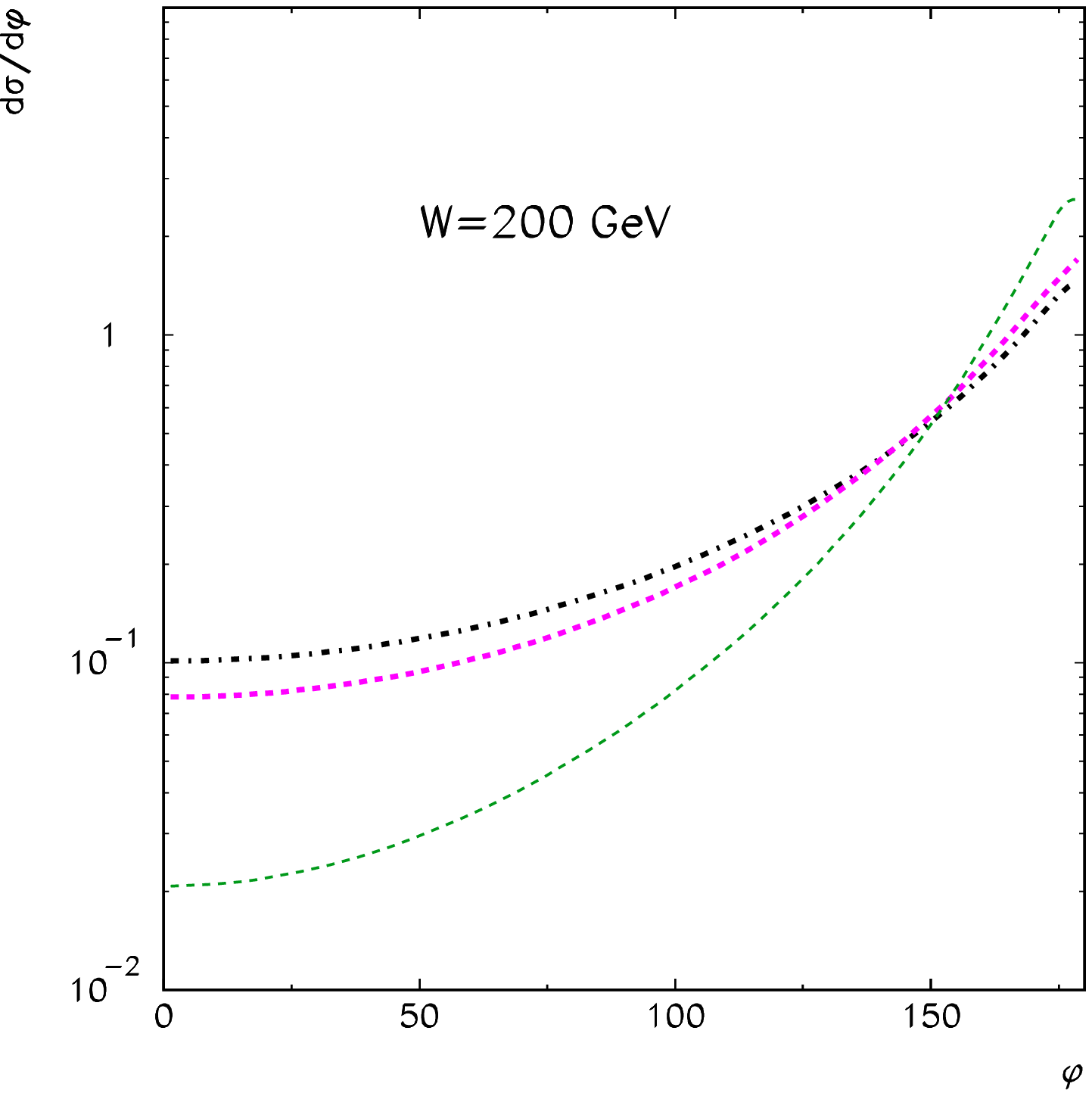}
\caption[*]{
Azimuthal correlations between $c$ and $\bar c$.
The theoretical results are compared to the recent results
from \cite{FOCUS} (fully reconstructed pairs).
\label{fig_phi}
}
\end{figure}

The azimuthal correlation functions $w(\phi)$
defined as:
\begin{equation}
w(\phi) =
 \int dp_{1,t}^2 \int dp_{2,t}^2 \int dz \;
\frac{f_g(x_g,\kappa^2)}{\kappa^4} \cdot
\tilde{\sigma}(W,\vec{p}_{1,t},\vec{p}_{2,t},z) \; .
\end{equation}
and normalized to unity
for two energies of $W$ = 18.4 GeV (FOCUS)
and $W$ = 200 GeV (HERA) are shown in Fig.\ref{fig_phi}.
The GBW-glue (thin dashed) gives too strong back-to-back correlations
for the lower energy. Another saturation model (KL, \cite{KL01})
provides more angular decorrelation, in better agreement with
the experimental data. The BFKL-glue (dash-dotted) provides very
good description of the data. The same is true for the
CCFM-glue (thick solid) and resummation-glue (thin solid).
The latter two models are more adequate for the lower energy.
In the present calculations we have used exponential form factor
with $b_e$ = 0.5 GeV$^{-1}$ (see \cite{KS04}).
For comparison in panel (b) we present predictions for $W$ = 200 GeV.
Except of the GBW model, there is only a small increase of
decorrelation when going from the lower fixed-order energy region
to the higher collider-energy region.

\section{Production of pions in NN collisions}

\begin{figure}[!thb]
\begin{center}
\includegraphics[width=3.2cm]{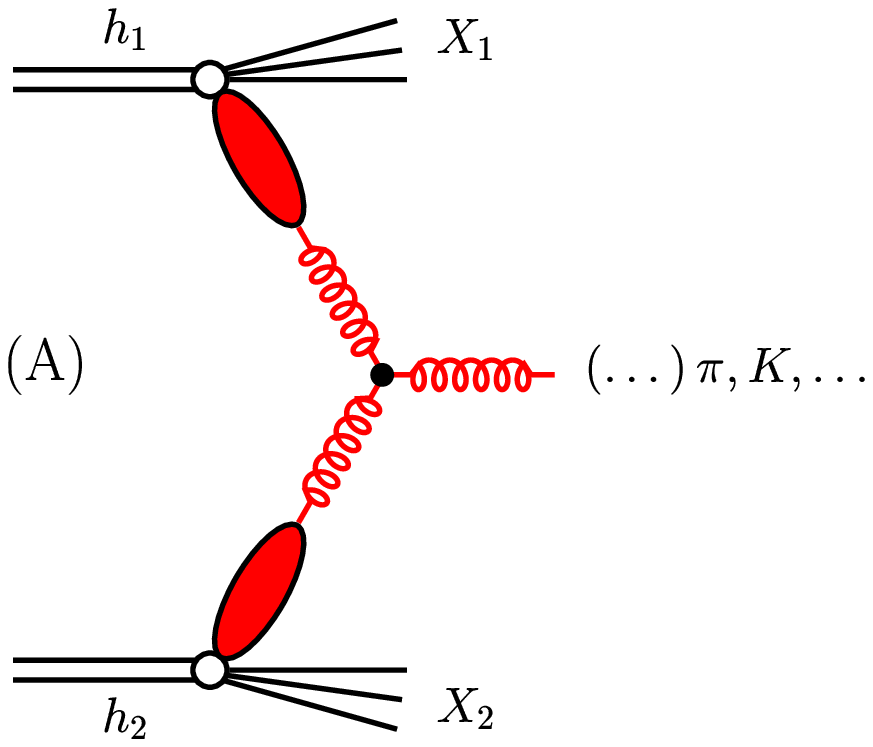}\\[30pt]
\includegraphics[width=3.2cm]{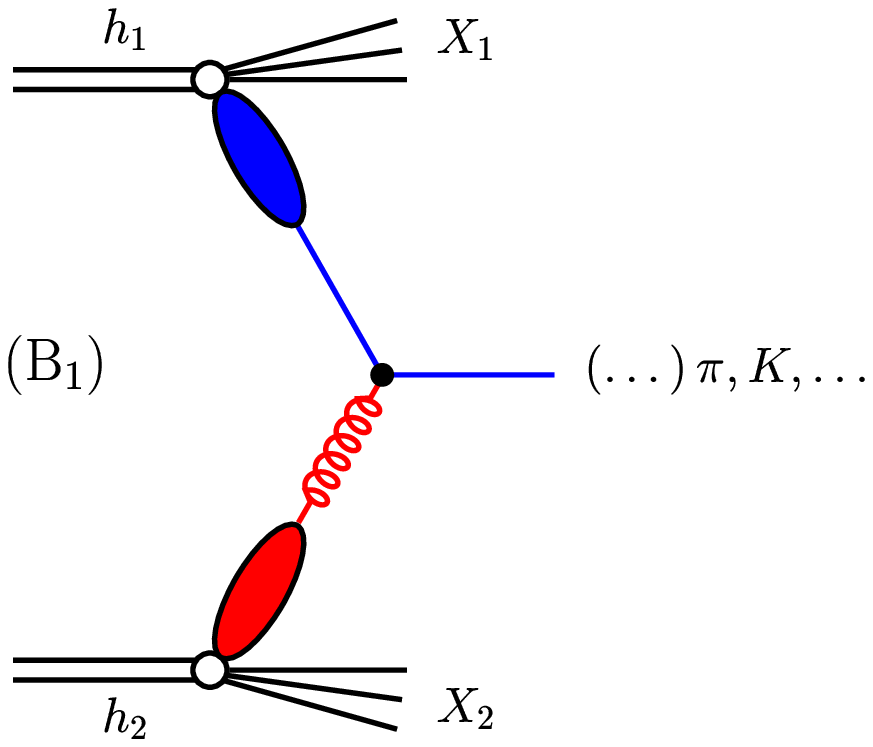}
\includegraphics[width=3.2cm]{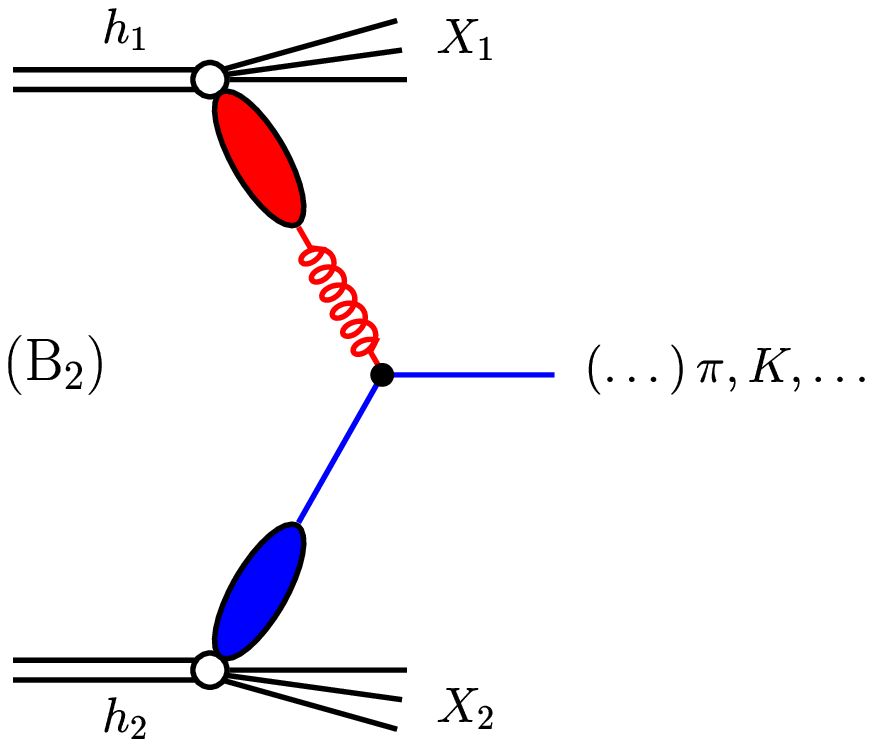}
\caption[*]{Leading-order diagrams for parton production
\label{fig:diagrams}
}
\end{center}
\end{figure}

\begin{figure} 
  \begin{center}
\includegraphics[width=6cm]{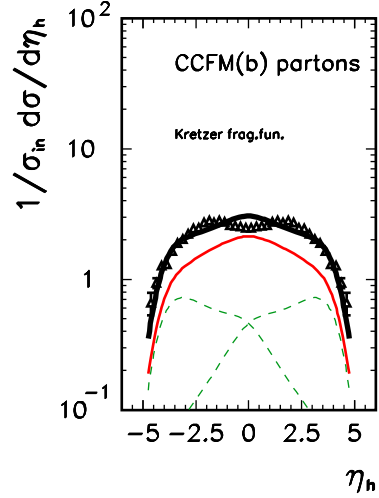}
  \end{center}
\caption[*]{
Pseudorapidity distribution of charged pions at W = 200 GeV
calculated with the CCFM parton distributions.
The experimental data are taken from \cite{UA5_exp}.
The thin solid line is the gluon-gluon contribution while
the dashed lines represent the gluon-(anti)quark
and (anti)quark-gluon contributions.
\label{fig:CCFM}
}
\end{figure}

The $gg \to g$ mechanism considered in the literature 
is not the only one possible.
In Fig.\ref{fig:diagrams} we show two other LO diagrams.
They are potentially important in the so-called
fragmentation region. The formulae for inclusive quark/antiquark
distributions are similar to formula for gluons and
will be given explicitly elsewhere \cite{SC05}.
The inclusive distributions of hadrons (pions, kaons, etc.)
are obtained through a convolution of inclusive distributions
of partons and flavour-dependent fragmentation functions
{\footnotesize
\begin{eqnarray}
\frac{d \sigma(\eta_h,p_{t,h})}{d \eta_h d^2 p_{t,h}} =
\int_{z_{min}}^{z_{max}} dz \frac{J^2}{z^2}  \nonumber \\
D_{g \rightarrow h}(z, \mu_D^2)
\frac{d \sigma_{g g \to g}^{A}(y_g,p_{t,g})}{d y_g d^2 p_{t,g}}
 \Bigg\vert_{y_g = \eta_h \atop p_{t,g} = J p_{t,h}/z}
 \nonumber \\
\sum_{f=-3}^3 D_{q_f \rightarrow h}(z, \mu_D^2)
\frac{d \sigma_{q_f g \to q_f}^{B_1}(y_{q_f},p_{t,q_f})}
{d y_{q_f} d^2 p_{t,q}}
 \Bigg\vert_{y_q = \eta_h \atop p_{t,q} = J p_{t,h}/z}
 \nonumber \\
\sum_{f=-3}^3 D_{q_f \rightarrow h}(z, \mu_D^2)
\frac{d \sigma_{g q_f \to q_f}^{B_2}(y_{q_f},p_{t,q_f})}
{d y_{q_f} d^2 p_{t,q}}
 \Bigg\vert_{y_q = \eta_h \atop p_{t,q} = J p_{t,h}/z}
 \; . \nonumber
\label{all_diagrams}
\end{eqnarray}
}
In Fig.\ref{fig:CCFM} we show the distribution in pseudorapidity
of charged pions calculated with the help of the CCFM parton
distributions \cite{GKB03} and the Gaussian form factor
(\ref{formfactor}) with $b_0$ = 0.5 GeV$^{-1}$, adjusted to
roughly describe the UA5 collaboration data.
Now both gluon-gluon and (anti)quark-gluon and gluon-(anti)quark
fussion processes can be included in one consistent framework.

As anticipated the missing up to now terms are more important
in the fragmentation region, although its contribution
in the central rapidity region is not negligible.
More details concerning the calculation will be presented elsewhere
\cite{SC05}.

\begin{figure}
\begin{center}
\includegraphics[width=6cm]{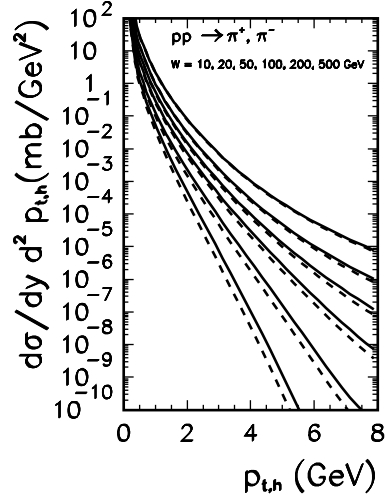}
\end{center}
\caption[*]{
Transverse momentum distribution of $\pi^+$ (solid) and $\pi^-$ (dashed)
mesons in proton-proton collisions for different center-of-mass energies.
\label{fig:w_pi}
}
\end{figure}

For completeness in Fig.\ref{fig:w_pi} we show transverse momentum
distribution of positive and negative pions for different incident
energies. The presence of diagrams $B_1$ and $B_2$ leads to
an asymmetry in $\pi^+$ and $\pi^-$ production. The higher
the incident energy the smaller the asymmetry. This is caused by
the dominance of diagram $A$ at high energies.

\section{Summary}

I have presented three examples of application of unintegrated parton
distributions to the description of three different processes.
Adjusting a value of one parameter of the nonperturbative form factor
a good description of different experimental data is achieved.
The uPDF which fulfil the Kwieci\'nski equations give a better
description of the intermediate-x data than other unintegrated
distributions in the literature. Many more additional tests are
possible and will be done in the future.

\end{document}